\documentclass[twocolumn,appendixfloats,tighten]{aastex6}
\usepackage{newtxtext,newtxmath}
\usepackage[T1]{fontenc}
\usepackage{ae,aecompl}

\usepackage{amsmath}	
\usepackage{amssymb}	

\usepackage{ctable}
\usepackage{url}
\usepackage{xspace}
\usepackage[normalem]{ulem}
\usepackage{hyperref}
\usepackage[all]{hypcap}
\usepackage{graphicx}

\def\msun{{\rm\,M_\odot}}

\def\msun{{\rm\,M_\odot}}

\newcommand{\kms}{\, {\rm km\, s}^{-1}}

\newcommand{\be}{\begin{equation}}
\newcommand{\ee}{\end{equation}}

\def\h2{${\rm\,H_2}$}

\usepackage{color}

\begin{document}

\title{Does GW190425 require an alternative formation pathway than a fast-merging channel?}

\author{Mohammadtaher Safarzadeh\altaffilmark{1}, Enrico Ramirez-Ruiz\altaffilmark{1,2}, Edo Berger\altaffilmark{3} }
\altaffiltext{1}{Department of Astronomy and Astrophysics, University of California, Santa Cruz, CA 95064, USA,
\href{mailto:msafarza@ucsc.edu}{msafarza@ucsc.edu}}
\altaffiltext{2}{Niels Bohr Institute, University of Copenhagen, Blegdamsvej 17, 2100 Copenhagen, Denmark}
\altaffiltext{3}{Center for Astrophysics | Harvard \& Smithsonian, 60 Garden Street, Cambridge, MA}

\begin{abstract}
The LIGO/Virgo Scientific Collaboration (LVC) recently announced the detection of a compact object binary merger, GW190425, with a total mass of $3.4^{+0.3}_{-0.1}$ M$_{\odot}$, and individual component masses in the range of about 1.1 to 2.5 $M_{\odot}$. 
If the constituent compact objects are neutron stars, then the total mass is five standard deviations higher than the mean of $2.66\pm 0.12$ M$_{\odot}$ for Galactic binary neutron stars (BNSs). 
LVC suggests such massive BNS systems are born from a fast-merging channel, and therefore, their non-detection in the Galaxy to be due to a selection effect. 
However, we are unable to reconcile the inferred formation efficiency from the reported merger rate, $\mathcal{R}_{\rm GW190425}=460^{+1050}_{-390}$ yr$^{-1}$ Gpc$^{-3}$, 
with predictions from our own study for fast-merging BNS systems.
Moreover, the comparable merger rates of GW190425, and GW170817 are possibly in tension with our results for two reasons: (i) more massive systems are expected to have a lower formation rate, and (ii) fast merging channels should constitute $\lesssim 10\%$ of the total BNS systems if case BB unstable mass transfer is permitted to take place as a formation pathway. We argue that to account for the high merger rate of GW190425 as a BNS system requires: (i) revisiting our understanding of NS formation in supernova explosions, or (ii): that more massive NSs need to be preferentially born with either very weak or very high magnetic fields so that they would be undetectable in the radio surveys. Perhaps massive NSs detected in NS-white dwarf binaries are our clues to the formation path of GW190425 systems.
\end{abstract}

\section{Introduction}
\label{sec:intro}
Recently, the LVC announced the detection of a compact object merger with a total mass of $3.4^{+0.3}_{-0.1}$ M$_{\odot}$, dubbed GW190425 \citep{Abbottetal:2020uq}. The merger's total mass lies in the range that each component can be consistent with being a neutron star. Prior to this detection, all the known binary neutron star (BNS) systems in the Galaxy occupied a very narrow range in mass, with a total mass of $\approx 2.66\pm 0.12 M_{\odot}$ \citep{Farrow:2019cl}, and the BNS system GW170817 with a total mass of $\approx 2.74^{+0.04}_{-0.01}M_{\odot}$ fell within this mass distribution as well \citep{Abbott:2017kt}.

If GW190425 is a BNS system, it is critical to explain why such a massive BNS system has not been observed in the Galaxy to date. One possible solution is the selection effects in the radio surveys. This can arise if such massive binaries are preferentially formed with very short periods, and thus they have inspiral phases below 10 Myr. This will subsequently lead to severe Doppler smearing \citep{Cameron:2018kj}, rendering their detection challenging. But does such a channel exist in nature?

BNS systems are thought to form through two distinct channels: Field formation \citep{Tauris:2017cf,Chruslinska:2018bv} and dynamical assembly \citep{RamirezRuiz:2015gl}, which is highly sub-dominant compared to field formation channel \citep{Ye:2020bv}. The field formation scenario predicts a delay time distribution that follows a power law \citep{Dominik:2012cw}. However, a sub-population of fast-merging systems can exist if unstable case BB mass transfer takes place in a common envelope between a He Hertzsprung gap (HG) star and a NS \citep{Ivanova:2003hc,Dewi:2003dy} depending on the component masses and orbital separation. The outcome of such unstable Case BB mass transfer is uncertain; these systems enter into Roche Lobe Overflow as a He HG stars, and as such, the donor stars in these systems lack clear core-envelope boundaries. Recent work by \citet{VignaGomez:2018eu} does not rule out the existence of an unstable Case BB phase when accounting for the Galactic BNS population, yet a stable case BB is still preferred. 

In this paper, we show that assuming GW190425 is a massive BNS system, the reported merger rate from LIGO/Virgo data is in tension with the expected merger rate of BNS systems born from fast-merging channels. In \S2 we estimate the merger rate of systems similar to GW190425, assuming they arise from a fast-merging channel. In \S3 we discuss the formation efficiency of such systems from population synthesis analysis. In \S4 we argue that if GW190425 is a BNS system, its high merger rate implies massive NSs are born with a rate similar to their lower mass counterparts but they are systematically dimmer and therefore undetectable in the radio surveys. This can happen if these systems have either very low or very high magnetic fields. We summarize the results in \S5.  

\section{Formation rate of GW190425 Type System as Fast merging BNS systems}

If such systems are the product of fast-merging channels, the estimate of their merger rate is simplified as the delay time distribution is short enough that it can be ignored \citep{Safarzadeh:2019kj}. The star formation rate of the universe can be parameterized as a function redshift as \citep{Madau:2014gta}:
\begin{equation}
\psi(z) = 0.015 \frac{(1+z)^{2.7}}{1+((1+z)/2.9)^{5.6}}\, \, M_{\odot}\, {\rm yr^{-1} \, Mpc^{-3}}.
\end{equation}
We combine this with the analytic expressions for metallicity evolution used from \citet{Eldridge:2019ev}. The fractional mass density of comoving star formation below metallicity $Z$ is given by:
\begin{equation}
\Psi\left(z,\frac{Z}{Z_{\odot}}\right)=\psi(z) \frac{\hat{\Gamma}[0.84,(Z/Z_{\odot})^2 \, 10^{0.3 z}]}{\Gamma(0.84)}.
\end{equation}
In this equation $\hat{\Gamma}$ and $\Gamma$ denote the incomplete and complete Gamma functions, respectively.  

The merger rate of such massive BNS systems is therefore given by:
\be
\mathcal{R}_{\rm GW190425} =\lambda_{\rm f,BNS} \Psi\left(0,Z\right) \times 10^9\,\, {\rm yr^{-1} \, Gpc^{-3}},
\ee
where $\lambda_{\rm f,BNS}$ is the formation efficiency that indicates the number of such systems born per unit solar mass of stars.

The reported value of $\mathcal{R}_{\rm GW190425}=460^{+1050}_{-390} {\rm yr^{-1} \, Gpc^{-3}}$ \citep{Abbottetal:2020uq} translates into a range of $\lambda_{\rm f,BNS}\approx 2.3\times 10^{-4} - 5\times10^{-3}$ if we assume that the progenitors of such systems have metallicities of $\lesssim 0.1 Z_{\odot}$ to have a total mass about the reported value \citep{Giacobbo:2018bw} due to fall-back mechanism and electron capture supernovae. If we relax the assumption that low metallicity is needed to produce such systems (i.e., that the efficiency of massive BNS formation from case BB unstable mass transfer is insensitive to progenitor metallicity) we find $\lambda_{\rm f,BNS} =7.5\times10^{-6}-1.6\times10^{-4}$. 

\section{Formation rate of fast-merging BNS systems in population synthesis models}

In \citet{Safarzadeh:2019dd} we analyzed formation models of BNSs from StarTrack \citep{Belczynski:2002gi,Belczynski:2006dq,Belczynski:2008kt} population synthesis code to search for fast merging channels. 
Three key parameters, and therefore eight different models were analyzed. 
The first parameter concerns with the behavior of binaries during a common envelope (CE) phase with a Hertzsprung gap (HG) donor star; during a CE, a NS enters the envelope of its companion, exchanging orbital energy to unbind the donor's envelope \citep{Fragos:2019ce} and accrete only modest amounts ($\lesssim 0.1 M_\odot$) of envelope material in the process \citep{MacLeod:2015bw}. 
For giant stars, with a clear core-envelope boundary, the end result of this process (so long as there is enough orbital energy available to keep the system from merging) is a closely bound binary comprised of the accretor star and the giant star's core. However, HG stars lack well-defined cores, and studies are inconclusive as to whether binaries entering into a CE during this phase can survive without merging \citep{Deloye:2010ey}.

In one submodel (A), HG stars are treated such that a core could be distinguished from an envelope in their evolutionary phase; hence a successful CE ejection is possible. In the other submodel (B), any system entering into a CE with an HG donor is assumed to merge. 

The second parameter studied was the natal kick received by a NS at birth. We analyzed models that adopted natal kicks randomly drawn from a Maxwellian distribution with $\sigma=265\kms$, based on the observed velocities of single Galactic pulsars \citep{Hobbs:2005be}, and models that adopt $\sigma=135\kms$. 

The third parameter we studied was metallicity where we explored two different metallicities: $Z=Z_{\odot}$ and $Z=0.1Z_{\odot}$ for the BNS populations. 

\begin{figure}
\includegraphics[width=\columnwidth]{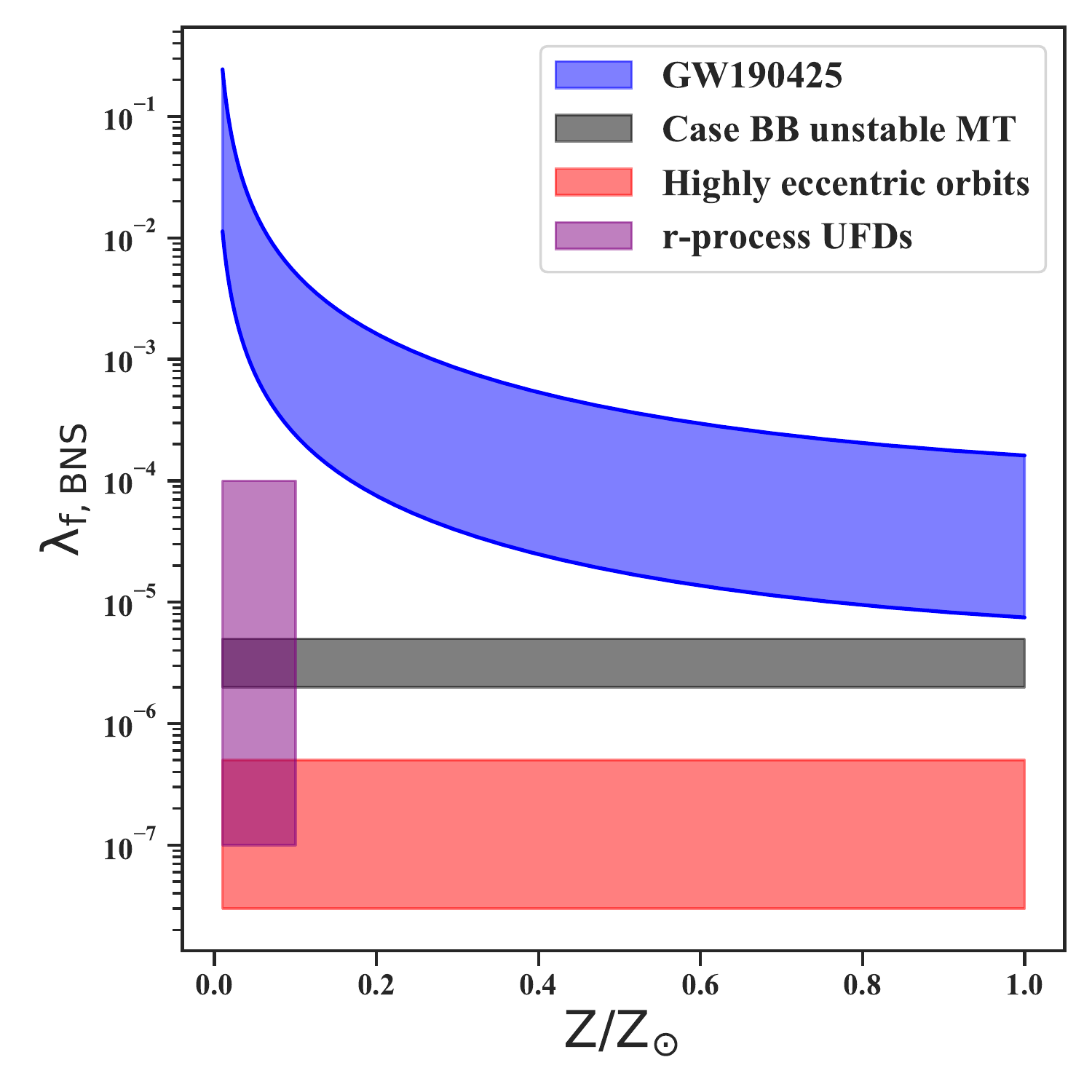}
\caption{The inferred formation efficiency of massive BNS systems if GW190425 represents such a system (blue shaded region) and formed from a fast-merging channel in the field. 
The x-axis shows maximum metallicity below which such a formation channel could be active. 
The black and red shaded regions represent the fast merging channel efficiency from population synthesis models. Two different sub-models, A and B have been analyzed, which treat the common envelope event with HG donor stars differently.
In sub-model A, the fast merging BNS population comes from systems experiencing case BB unstable mass transfer (MT). In sub-model B, the fast merging channels are highly eccentric BNSs. 
The uncertainty in each band comes from different assumptions regarding the natal kicks of the neutron stars at birth and the metallicity dependence of the efficiency of the CE phase. 
The large formation efficiency required to account for GW190425 as a BNS system born out of a fast merging channel is inconsistent with the formation efficiencies expected in the population synthesis models \citep{Dominik:2012cw}. At low metallicities, the fast merging BNS mergers statistics should be compared to that of r-process enriched ultra-faint dwarf galaxies which we show with a purple shaded region.}
\label{fig:1}
\end{figure}

In submodel A, case BB unstable mass transfer make up a population of fast merging channel BNSs. In submodel B, the fast merging channel comes from BNSs on highly eccentric orbits due to natal kicks in favorable directions. The summary of the analysis is presented in Figure \ref{fig:1}. 

The formation rate of fast-merging BNS systems is about $2-5\times10^{-6}M_{\odot}^{-1}$ for submodel A where the range spans the uncertainty to the metallicity and natal kick assumptions. For submodel B the formation efficiency is about $3\times10^{-8}-5\times10^{-7}$. Such small formation channels are inconsistent with the required large formation efficiency of $\lambda_{\rm f,BNS} = 2.3\times10^{-4}- 5\times10^{-3}$ inferred from GW190425 if this system is a BNS system formed through a fast-merging channel at low metallicities. Even if we treat the metallicity as a parameter with unknown impact, the absolute minimum efficiency of $\lambda_{\rm f,BNS} =7.5\times10^{-6}$ is still in tension with the efficiency range in population synthesis models of $\lambda_{\rm f,BNS}=2-5\times10^{-6} M_{\odot}^{-1}$.

However, we note that there is considerable uncertainty involved in population synthesis models, and one can not, in the end, rule out whether GW190425 was formed through an unstable case BB mass transfer. 
For example, \citep{RomeroShaw:2020wa} suggests this as a formation scenario for GW190425, arguing that the helium progenitor star could explain the high mass of the GW190425. 
While we have independently confirmed our results using COSMIC \footnote{https://cosmic-popsynth.github.io} population synthesis model, we note that other groups have hinted on low rates of BNS merger in the local universe, in general: 
For example, through a systematic study, \citet{VignaGomez:2018eu} shows that a scenario in which case BB unstable mass transfer is modeled in their population synthesis code can lead to high local merger rates. However, one has to note that i) such a model would be in tension with the distribution of the Galactic BNS systems (see their Figure A1), meaning not all the BNSs could experience this channel, and ii) the study by \citet{VignaGomez:2018eu} is carried out at $Z=0.0142$, and local star formation at such low metallicities is suppressed. Therefore, the predicted rates from this channel will be suppressed. Moreover, in a separate study, \citet{Neijssel:2019ca} takes into account the uncertainty in the metallicity and star formation history of the universe and find generally low local merger rate for BNS systems ranging from 10-300 ${\rm yr^{-1} \, Gpc^{-3}}$ (see their Table C.1). Therefore, if the rates coming from population synthesis are low, any sub-set of BNSs, e.g., a system similar to GW190525, will be predicted to have low merger rates as well.

Separately, models attempting to explain GW 190425 as a fast merging system need to be tested against the rarity of r-process enriched ultra-faint dwarf galaxies (UFDs): if the BNS merger rates from a channel active at low metallicities is hight, it can lead to r-process enrichment of UFDs. However, of all the UFDs in the Galaxy, only about 20\% are r-process enriched \citep{Safarzadeh:2019dd}. Through Poisson statistics of r-process sources in the early universe, \citet{Beniamini:2016bf} showed the formation efficiency per solar mass of stars formed should be within $10^{-7}-10^{-4}\msun$. Based on our model, such formation efficiency at low metallicities falls short of explaining GW190425, since star formation rate at such low metallicities is suppressed in the local universe.
We show the expected efficiency of fast-merging BNS system through the statistics of r-process enriched UFDs from \citet{Beniamini:2016bf} in Figure \ref{fig:1}.
 
We emphasize that the argument based on the rate of case BB unstable mass transfer presented in this work, although does not refute this formation path, casts doubt on its efficiency to explain GW190425, and further detailed work would be needed in support or against our claim.

\section{Possible solutions}
Our goal in this paper is not providing a progenitor solution but instead testing whether or not the presence of GW190425 can be accommodated within our current understanding. 
Here we discuss some plausible alternatives. Each possibility would require further calculations and simulations to be effectively tested, which is beyond the scope of this paper.

\begin{figure}
\includegraphics[width=\columnwidth]{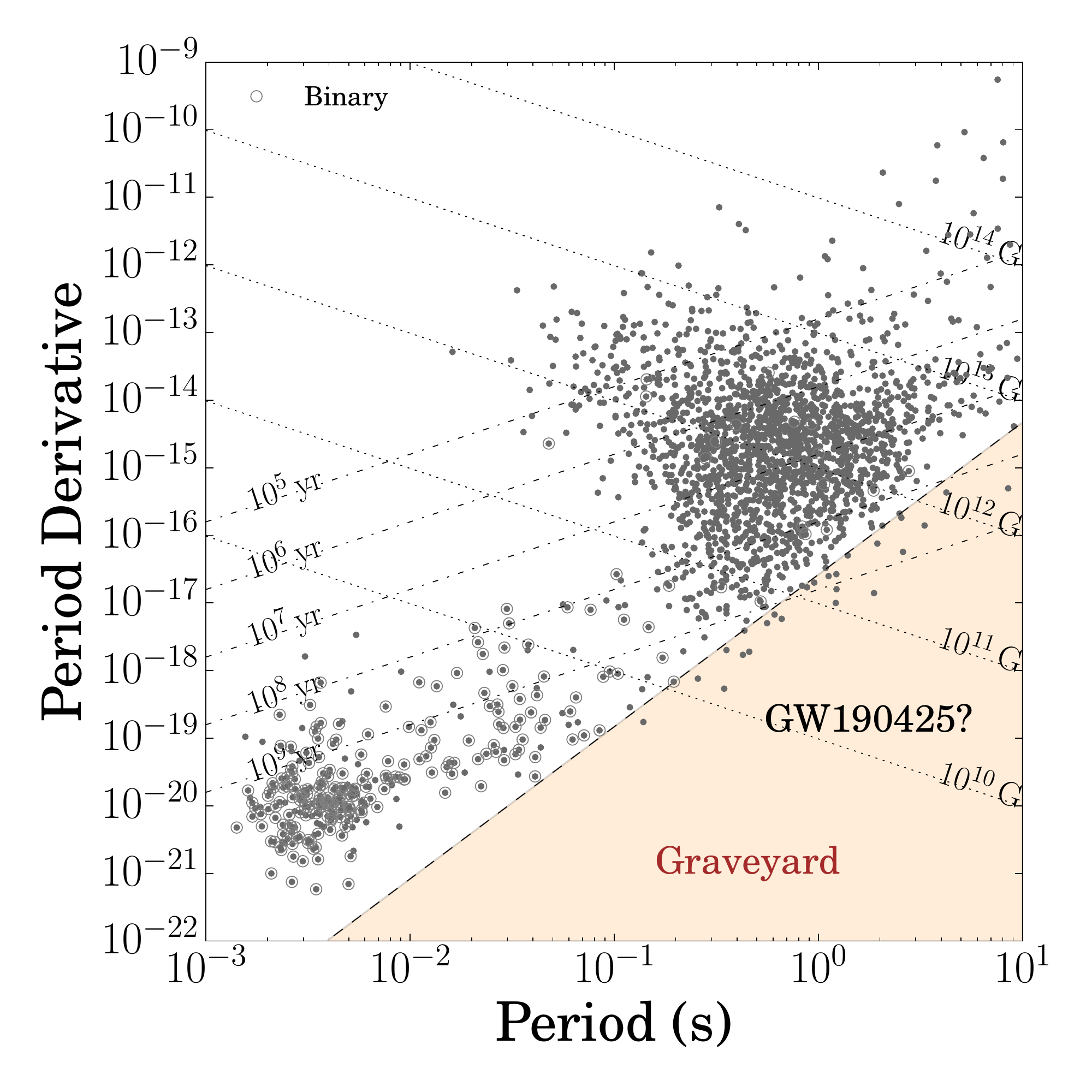}
\caption{The distribution of known pulsars in period-period derivative plane.  Pulsars in binaries are shown with a circle and GW170817 is argued to be consistent with this population \citep{RamirezRuiz:2019jp}.
The graveyard is the region shaded with the brown color, where pulsars would not be detectable in radio sky. A NS can either be born in the graveyard or migrate there due to the spin down effect of its magnetic dipole moment. The timescale for the spin down can be very short for those with high magnetic fields.
Since no binary neutron star is observed to be as massive as GW190425, it is possible that such massive binaries merge while being in the graveyard. 
This plot is made using the ATNF pulsar catalog \citep{Manchester:2005hq}, and PSRQPY Python package \citep{Pitkin:2018tx}.}
\label{fig:p-pdot}
\end{figure}

The lesson we learn from GW190425 is that nature makes BNS systems that are i) massive and ii) currently not observable in the radio sky as pulsars.
These two points could be related. Such systems could reside in the graveyard parameter space in the period-period derivative distribution of known pulsars (see Figure \ref{fig:p-pdot} ).

We discuss possible scenarios which can produce massive BNS systems that have either a very weak or very high magnetic field dipole moments, which will make them undetectable. 
 
\subsection{The birth magnetic fields of neutron stars}
Neutron stars are expected to rotate extremely rapidly and have strong magnetic fields due to the conservation of magnetic flux. The observation of several neutron stars, known as central compact objects \citep[CCOs; ][]{Ho:2012ct}, in supernova remnants with weaker magnetic fields than the average radio-pulsar population has motivated a lively debate about their formation. Can GW190425 be associated with this population?

The reported merger rate of double neutron stars from the LIGO VIRGO collaboration, $\mathcal{R}_{\rm GW190425}=460^{+1050}_{-390} {\rm yr^{-1} \, Gpc^{-3}}$, $\mathcal{R}_{\rm GW170817}=760^{+1740}_{-650} {\rm yr^{-1} \, Gpc^{-3}}$, is similar for the massive and light components.

In a recent study \citet{Sukhbold:2016bo} showed that a small percentage ($\lesssim 10\%$) of stars with a mass between $18-120~\msun$ could make NSs with mass greater than $1.6~\msun$, while NSs with mass consistent with the observed galactic population could form from stars with a mass between $10-18~\msun$. Assuming a Salpeter initial mass function following $dN/dM\propto M^{-2.35}$, the relative expected formation rate of GW190425 to GW170817 type systems should be at least (conservatively) less than $10\%$. 
From this, we can conclude that the progenitor masses of these systems can not be very different since the initial mass function of stars would have drastically impacted the expected merger rates. 
We note our arguments are based on the inferred merger rate from a single detection of GW190425-like systems. 
If future LVC observations show that the merger rate of GW190425-like systems is in fact smaller than GW170817-like systems, the above stated tension would be reduced.

One possibility is that the formation rate of heavier NSs is similar to standard NS formation. In order for these systems to be undetected, we would require that heavier NSs are born preferentially either inside the graveyard, such as CCOs, or are born with extremely large magnetic fields such that their radio pulsar lifetime becomes extremely short. 

CCOs might appear to be a natural candidate as they are likely NSs born with weak magnetic fields \citep{Ho:2012ct}, although fallback of the supernova debris onto the neutron star could have been responsible for the submergence of the field and its apparently low value. In such cases, re-surfacing and emergence of the field could make these objects detectable. If born with a weak field, CCOs can only work as progenitors if they are more massive at birth than pulsars. If these systems are not born with higher masses, post-birth mass accretion is likely to spin them up to high, detectable luminosity, as we argue in the following section. Many issues about the nature of CCOs need to be settled before we can derive any firm conclusions.  

\subsection{Consequences of significant mass gain}
Mass accretion can take place during common envelope evolution, although estimates for the total mass accreted have been shown to be small \citep{MacLeod:2015bw}.
Stable mass transfer through Roche Lobe overFlow is another possibility. Yet, a NS accreting at the Eddington limit, $\dot{M}_{\rm Edd}\lesssim 10^{-8} M_\odot$/yr can accrete less than 0.1 $M_\odot$ in about 10 Myrs, which is a conservative estimate for the lifetime of the companion. Assuming both neutron stars have 1.4 solar mass at birth, the first neutron star needs to accrete about 0.6 $\msun$ to bring the total mass of the binary to about 3.4 $\msun$. The accretion of this material will naturally lead to the NS spinning up and, as a result, the object luminosity will undeniably increase (Figure~\ref{fig:p-pdot}). If the magnetic field of the NS is large, it is feasible for the spin down timescale to be short, which might account for its potential disappearance.  

Another opportunity to grow is by accreting fallback material from the supernova that gave rise to the last NS. The distribution of orbital separations for the pre-SN progenitor and the NS are poorly known \citep{Ivanova:2003hc} and it can range from hours to few days.
If the SN explosion is not as energetic, one can expect to accrete material that accumulates in a circumbinary disk \citep[e.g.,][]{Schroder:2018ko}. In order for the NS to accrete 0.6 $\msun$ of fallback material, two properties need to be satisfied: The binary needs to be compact and the energy injected in the supernova needs to be no more than a few times the binding energy of the exploding He star, as demonstrated by \citet{Schroder:2018ko}.  

This might happen only in a few cases, thus making it potentially difficult to still reconcile the merger rates of heavy systems with those from the standard channel.

One consequence of this super Eddington mass accretion phase would be the subsequent burying of the magnetic field of the NS \citep{Bernal:2013br}, yet the field emergence timescale is likely to be significantly shorter than the merging timescale and the NSs will likely spin up as a result of accretion. Both effects will help increase the pulsar luminosity. As we argued before, only a very low or very high magnetic NS will be able to appear undetected in our Galaxy if this channel was a prominent one.  

In the end, the key unresolved issue is that we have serious deficiencies in our understanding of the magnetic field evolution and potential decay in pulsars. 
Past research attempting to constrain such initial distributions relied on the observed pulsar distribution in period-period derivative phase space in the Galaxy \citep{FaucherGiguere:2005hn,Osiowski:2011hi}. 
In this scenario the distribution of NSs in the $p-\dot{p}$ phase depends sensitively on assumptions 
relating to the decay of the magnetic field \citep{Romani:1990kg,Cumming:2001jg,Melatos:2001fx,Choudhuri:2002df,Payne:2004er}. this is because the surface magnetic field strength of millisecond pulsars is found to be about 4 orders of magnitude lower than that of garden variety radio pulsars (with a spin of $\approx0.5-5$ s and $B\approx10^{12}$G). Without a clear understanding of this, we would be unable to make detailed predictions about the detectability of GW190425-like systems in the Galaxy. 
 
\section{Summary \& Discussion}

The LVC recently announced the detection of a compact object binary merger, GW190425, with a total mass of $3.4^{+0.3}_{-0.1}$ M$_{\odot}$. This system lies five standard deviations away from the known Galactic population of binary neutron stars (BNSs) with a mean total mass of $2.66^{+0.12}_{-0.12} M_{\odot}$. The comparable merger rate of this system to GW170817 raises several issues that we attempted to elucidate in this work.

The LVC speculates that such massive system were not detected in previous radio surveys due to selection effects pertinent to far-merging channel BSNs.
Assuming such systems are born from a fast-merging channel, namely, in a case BB unstable mass transfer, it would indicate that the delay time of such a system is extremely short (less than 10 Myr). To be consistent with the reported merger rate of GW190425 from LIGO O3 data ($\mathcal{R}_{\rm GW190425}=460^{+1050}_{-390} {\rm yr^{-1} \, Gpc^{-3}}$), one concludes that the efficiency of formation of fast merging BNS systems should be between $\lambda_{\rm f,BNS}\approx2\times10^{-4}-5\times10^{-3} M_{\odot}^{-1}$ depending on what we assume as the maximum allowable metallicity of their progenitor stars. However, we are unable to reconcile this with the formation efficiency of fast merging BNS systems from the population synthesis models studied in this work which point to $\lambda_{\rm f,BNS}=2-5\times10^{-6} M_{\odot}^{-1}$.

Moreover, the comparable merger rate challenges our understanding of supernova explosion in massive stars as more massive NSs are born from heavier progenitors such that the relative formation rate of massive to normal BNS systems should be at least suppressed by an order of magnitude. 

Regardless of the issues above, if we assume our understanding of the supernova and fall back physics is subject to drastic modifications, we suggest that the only way to reconcile the observed rate with the lack of the previous detection of such systems in radio surveys is if these systems have suppressed magnetic dipole moment. 
We argue a plausible solution could be that more massive NSs need to be preferentially born with either very weak or very high magnetic fields so that they would be undetectable in the radio surveys. 

However, a caveat to this hypothesis is that we have seen massive NSs in white dwarf-neutron star (WD-NS) binaries \citep{Kiziltan:2013ky}, such as J0740+6620. 
This means massive NSs are born with normal magnetic field strength in WD-NS systems and if our theory holds, it should be able to account for this.
Here we note that a WD-NS systems can have different formation pathways, involving many episodes of mass transfer \citep{Toonen:2018bt} such that is is the WD that is born first, and then the NS. 
Extension of this work to WD-Ns would require modeling the magnetic field evolution in population synthesis studies of such systems which is beyond the scope of our work.

\acknowledgements We thank the referee for their instructive comments. We are also thankful to Alejandro Vigna-Gomez for insightful discussions. MTS is grateful to the Center for Astrophysics for hospitality during this work. ER-R and MTS thank the Heising-Simons Foundation, the Danish National Research Foundation (DNRF132), and NSF (AST-1911206 and AST-1852393) for support. EB acknowledges support from NSF grant AST-1714498. We are thankful to Ryan Foley, Duncan Lorimer, and Abraham Loeb for thoughtful discussions. 

\bibliographystyle{yahapj}
\bibliography{the_entire_lib.bib}
\end{document}